\documentstyle[twocolumn,amssymb,aps,psfig,floats,amsmath]{revtex}

\begin{document}
\tolerance 50000

\draft

\twocolumn[\hsize\textwidth\columnwidth\hsize\csname @twocolumnfalse\endcsname

\title{Dynamical properties of low dimensional CuGeO$_3$ and NaV$_2$O$_5$ 
spin-Peierls systems}

\author{David Augier and Didier Poilblanc }
\address{
Laboratoire de Physique Quantique \& Unit\'e Mixte de Recherche 
CNRS 5626\\ 
Universit\'e Paul Sabatier, 31062 Toulouse, France.}

\date{July 97}
\maketitle

\begin{abstract}
\begin{center}
\parbox{14cm}{

Properties of low-dimensional spin-Peierls systems
are described by using a one dimensional $S=\frac{1}{2}$ 
antiferromagnetic Heisenberg chain 
linearly coupled to a single phonon mode of wave vector $\pi$ (whose
contribution is expected to be dominant). 
By exact diagonalizations of small rings with up to $24$ sites
supplemented by a finite size scaling analysis,
static and dynamical properties are investigated.
Numerical evidences are given for a spontaneous discrete symmetry breaking 
towards a spin gapped phase with a frozen lattice 
dimerization. Special emphasis is put on the comparative study of 
the two inorganic spin-Peierls compounds CuGeO$_3$ 
and NaV$_2$O$_5$ and the model parameters are determined from a fit of 
the experimental spin gaps. 
We predict that the spin-phonon coupling is 2 or 3 times
larger in NaV$_2$O$_5$ than in CuGeO$_3$.
Inelastic neutron scattering spectra are calculated and 
similar results are found in the single phonon mode approximation
and in the model including a static dimerization.
In particular, the magnon $S=1$ branch is clearly
separated from the continuum of triplet excitations by a finite gap.}
\end{center}
\end{abstract}

\pacs{
\hspace{1.9cm}
PACS numbers: 64.70.Kb, 71.27.+a, 75.10.Jm, 75.40.Mg, 75.50.Ee}
\vskip2pc]

\section{Introduction}
\label{I}
Recently, a renewed interest for one dimensional (1D)
spin chains was created by the observation of 
spin-Peierls transitions in the inorganic compounds 
CuGeO$_3$~\cite{hase} and 
$\alpha'$-NaV$_2$O$_5$~\cite{isobe,fujii,ohama,weiden}. Below 
some critical temperature $T_{\mathrm{SP}}$,
the spin-Peierls phase is experimentally inferred from a
rapid drop of
the spin susceptibility. The low temperature phase
is characterized by the opening of a spin
gap $\Delta$ (see below) and the dimerization of the lattice along the 
chain direction as confirmed for example 
by X-rays diffraction in CuGeO$_3$~\cite{pouget} and 
NaV$_2$O$_5$~\cite{fujii} or by Na NMR 
experiments in NaV$_2$O$_5$~\cite{ohama}.

In general, these compounds are
well described above the transition temperature by a 
1D frustrated antiferromagnetic (AF) Heisenberg chain.
The nearest neighbor and next-nearest neighbor spin exchange 
integrals $J$ and $J'$ 
can be determined by a fit
of the magnetic susceptibility $\chi$ at high temperatures. 
In fact the position of the maximum of the curve and more generally the 
magnetic properties only depend on 
the frustration ratio $\alpha=J'/J$.
 
Parameters such as
$J=160\ K$ and $\alpha=0.36$ \cite{rieradobry} or 
$J=150\ K$ and $\alpha=0.24$ \cite{castilla} have been suggested
for CuGeO$_3$. Further studies~\cite{rierakoval,bouzerar} seem
to confirm that the dimerization is large in this system and
we shall take $\alpha=0.36$ in the rest of the paper. For the novel 
NaV$_2$O$_5$ system $J=440\ K$ and
$\alpha\approx0$ have been proposed~\cite{weiden} in good
agreement with Refs.~\cite{isobe,mila}. 

The zero temperature spin gap $\Delta$ has been
determined by several means.
Inelastic neutron scattering (INS) gives a direct measure of it.
So far INS has been performed on single crystals of 
CuGeO$_3$ and powder NaV$_2$O$_5$ samples.
Values of $\Delta\simeq2.1$ meV~\cite{nishi,regnault}
and $\Delta=9.8$~meV~\cite{fujii} have been reported,
respectively. Other more indirect methods like NMR can also provide 
a measure of the magnitude of the spin gap. 
$^{63}$Cu or $^{65}$Cu NMR have been performed on single crystals of 
CuGeO$_3$~\cite{itoh} and $^{23}$Na 
NMR on aligned polycrystals of  
NaV$_2$O$_5$~\cite{ohama}. The local 
magnetic susceptibility is proportional to the NMR Knight shift and the 
spin gap is estimated by a fit of the temperature dependence of the 
local susceptibility below the transition temperature. A value of
$\Delta\simeq 8.4$~meV was found for NaV$_2$O$_5$. 
A third independent estimation of $\Delta$ can be obtained by
a fit of the low temperature bulk magnetic susceptibility of single crystals
measured for instance by a SQUID-magnetometer technique.
Ref.~\cite{weiden} reports
a value of $\Delta\simeq7.3$~meV for NaV$_2$O$_5$. 

The dimensionless ratio $\Delta/J$ is the crucial parameter needed
in our theoretical analysis. Results for CuGeO$_3$ are now well established
and the value of $\Delta/J=0.151$ is often used in the literature. 
Nonetheless ratios such as $\Delta/J=0.203$~\cite{fujii},
$\Delta/J=0.175$~\cite{ohama} or 
$\Delta/J=0.193$~\cite{weiden} can be found for 
NaV$_2$O$_5$. Since this last estimation
was obtained from experiments performed on single 
crystals we have thus decided to use it as a reference. In any case,
the quite small differences between the previous experimental 
values are not relevant.

Theoretically, the spin dynamics of the 1D Heisenberg chain depends 
strongly on the frustration parameter $\alpha$. 
Indeed, for $\alpha>\alpha_c\simeq0.241$ a gap appears in the 
spin excitation spectrum~\cite{haldane,okamoto}.
Therefore, we expect that the two previous spin-Peierls compounds 
will have quite different magnetic properties. CuGeO$_3$ is dominated by 
intrachain frustration. On the other hand,
NaV$_2$O$_5$ will behave, at high temperatures,
more closely to an unfrustrated Heisenberg chain and, at low temperatures,
the small interchain frustration alone cannot be responsible 
for the opening of a spin gap. The coupling
to the lattice is therefore expected to play a dominant role
in the transition at least for NaV$_2$O$_5$. 
In order to study their interplay, the frustration and the spin-lattice
coupling have to be treated on equal footings. This is the purpose of
this paper.

It is well known that a $1D$ system 
shows no phase transition at finite temperature because
of quantum fluctuations. Interchain couplings are necessary to obtain
a finite transition temperature. However, they are thought to be small and
will be neglected hereafter in the study of zero temperature 
properties.

So far, there have been various
attempts to treat the coupling to the lattice
by considering a static dimerization $\delta$ of the exchange integral
(so called adiabatic approximation or frozen phonon approximation).
The value of $\delta$ is determined in order 
to obtain the experimental value of the zero temperature 
spin gap $\Delta$. Dimerizations such as $\delta=0.014$~\cite{rieradobry}
and $\delta=0.048$~\cite{augier} 
were proposed for CuGeO$_3$ and for NaV$_2$O$_5$, 
respectively, in order to reproduce the measured spin gaps 
(assuming $\Delta\simeq0.151J$ and $\Delta\simeq0.193J$ 
for CuGeO$_3$ and NaV$_2$O$_5$, respectively).
Calculations using this approach have been performed 
in order to make first comparisons with 
experiments~\cite{rierakoval,haas,poilblanc,muth}. 

In this paper we use a modification of the previous static model
to describe the physical properties of one-dimensional
spin-Peierls compounds below the transition temperature.
For convenience, the previous {\em ad hoc} static 
dimerization discussed above is replaced here by 
a single dynamical optical phonon mode (Section~\ref{II}).
As far as thermodynamic properties are concerned,
this model should be, in fact, equivalent {\it in the thermodynamic limit} 
to a model where the lattice is treated at a mean field level~\cite{lieb}.
However, this new approach has some advantages:
(i) it incorporates automatically the elastic energy and avoids the
lengthy iterative procedure needed in a mean-field treatment to converge to
the equilibrium static lattice dimerization; (ii) it enables to study
the mechanism of the lattice symmetry breaking and, hence, provides a
basis for future studies including a macroscopic number of phonon
modes (i.e. proportional to the system length $L$)~\cite{wellein} in
spin-Peierls chains. 

Within this single mode approximation, we
truncate the Hilbert space of the phonons and show in details that
this approximation is well controlled (Section~\ref{III}). Using a 
finite size scaling analysis (discussed in detail in Section~\ref{IV})
the dimerization and the spin gap resulting from a spontaneous
discrete symmetry breaking of the lattice periodicity are
calculated (Section ~\ref{V}). 
Focussing primarily on CuGeO$_3$ and NaV$_2$O$_5$ materials,
we then establish 
a simple relation between the parameters of the model in such a way to enforce
the constraint that the numerically calculated
spin gap is equal to the experimental gap. 
The role of the parameters is discussed.
In the last part (Section~\ref{VI}), we study the spin dynamics.
In particular, we investigate the role of the 
lattice dynamics on the low energy magnon branch and low energy structures
in the dynamical spin structure factor.
Our results are compared to the ones obtained in the static
model~\cite{augier,poilblanc,uhrig,fledder}.

\section{Models}
\label{II}
Our starting point is the 1D frustrated AF Heisenberg chain.
For practical applications, the previous values of $J$ and $\alpha$ will
be used~:
$J=160\ K$, $\alpha=0.36$ for CuGeO$_3$~\cite{rieradobry} and 
$J=440\ K$, $\alpha=0$ for NaV$_2$O$_5$~\cite{weiden}.
In addition, a coupling between spins and
dispersionless optical phonons (magneto-elastic coupling) is considered. 
For sake of simplicity we assume a linear dependence of the
exchange integrals on the relative atomic displacements 
$\{u_i\}$~\cite{voit,khomskii},
\begin{equation}
H=J \sum_i ( (1+\lambda u_i)\; \vec{S}_{i}.\vec{S}_{i+1}+
\alpha \vec{S}_{i}.\vec{S}_{i+2} ) +H^0_{\mathrm{ph}},
\label{debut}
\end{equation}
where $\lambda$ is the coupling constant. $H_{\mathrm{ph}}^0$ is 
the phononic Hamiltonian of identical independent quantum oscillators,
\hbox{$H^0_{\mathrm{ph}}=\sum_{i}(\frac{p_i^2}{2m}+\frac{1}{2}K\,u_i^2)$}
($p_i$ is the conjugate momentum associated to 
the atomic displacement $u_i$).
The  atomic displacements  $u_i$ and their conjugate variables can easily
be expressed in term of the canonical phonon creation and annihilation
operators $b^\dagger_k$ and $b_k$.
Since the spin susceptibility diverges 
(for $\alpha<0.5$~\cite{tonegawa,chitra}) at 
momentum $k=\pi$ we expect that the coupling to the lattice 
will be dominant at $k=\pi$ which corresponds, in fact, to the 
modulation of the spin-Peierls ground state.
Therefore, from now on, we shall only keep a 
single $k=\pi$ phonon mode~\cite{dobry}. In this case, using,
$$u_i\simeq(-1)^i \sqrt{\frac{1}{2mL\Omega}} (b^{\phantom\dagger}
_{\pi}+b^{\dagger}_{\pi})$$
($\Omega ^2=K/m$ and $L$ is the number
of sites), the final Hamiltonian becomes,
\begin{eqnarray}
H\!\!=\!\!J\! \sum_i \big\{
\big(1+ &\frac{g\,(-1)^i}{\sqrt{L}} (^{\phantom\dagger}
b_{\pi}+b_{\pi}^{\dagger})\big) \;
\vec{S}_{i}.\vec{S}_{i+1}+
\alpha \vec{S}_{i}.\vec{S}_{i+2} \big\}  \cr
&+H^{0}_{\mathrm{ph}}, 
\label{hamildyn}
\end{eqnarray}
where $g=\lambda\sqrt{\frac{1}{2m\Omega}}$ is the dimensionless 
coupling constant. Within this approximation 
$H^0_{\mathrm{ph}}$ can be rewritten as \hbox{$H^0_{\mathrm{ph}}=
\Omega\ (b_{\pi}^{\dagger}b^{\phantom\dagger}_{\pi}+\frac{1}{2})$}
where $\Omega$ is the energy of a phononic quantum. 

Before going further, we can already discuss qualitatively 
the physics contained in Hamiltonian~(\ref{hamildyn}).
Indeed, we expect in the thermodynamic limit a discrete symmetry breaking
corresponding to a doubling of the unit-cell. 
This can be described very simply at the MF level.
By assuming a dimerization $\delta$ given by the 
order parameter $\frac{g}{\sqrt{L}}\langle
b_{\pi}^{\phantom\dagger}+b^{\dagger}
_{\pi}\rangle_{MF}$ and omitting a constant part, the MF
Hamiltonian takes the form,
\begin{eqnarray}
H_{\mathrm{MF}}=J\sum_i ((1&+\delta(-1)^i)  \vec{S}_{i}.\vec{S}_{i+1}
+\alpha \vec{S}_{i}.\vec{S}_{i+2}) \cr
& +\frac{1}{2}L\frac{K}{\lambda^2}\delta^2,
\label{hamilstat}
\end{eqnarray}
where the last term is the elastic energy loss. This is 
exactly the well known model
describing a static dimerization below the transition temperature 
in spin-Peierls systems~\cite{rieradobry,castilla}. 
Interestingly enough, a similar effective model has also been used
to describe conjugated hydrocarbons with bond alternation 
such as polyacetylene~\cite{malrieu}.
In this new form, the breaking of the lattice periodicity is explicit. 
As a consequence the ground state becomes doubly 
degenerate (the order parameter $\delta$ can take a positive or
a negative value) and a spin gap appears. 
The spin-Peierls ground state is characterized by a ``$\cdots A-B-A-B
\cdots$'' 
pattern with a succession of strong singlet A bonds and weak singlet B bonds
(so called Valence Bond or dimer state). 
Note that $\delta$ in model (\ref{hamilstat}) is a variational parameter
to be determined in order to minimize the ground state energy by 
an iterative procedure. In contrast, the dimerization in Hamiltonian 
(\ref{hamildyn}) arises from a dynamical symmetry breaking.
However, it is interesting to notice that 
models (\ref{hamilstat}) and (\ref{hamildyn}) should be in fact 
equivalent~\cite{notte} 
 in the {\it thermodynamic limit}, at least as far as their 
thermodynamic properties are concerned~\cite{lieb,note1}. 

Static and dynamical quantities are given by exact diagonalizations of 
small chains. Using a finite size scaling analysis, results in the
thermodynamic limit are deduced.  
The parameters $\delta$ on one hand and 
$g$ and $\Omega/J$ on the other hand are determined from a fit to the
experimental spin gap. 

\section{Truncation procedure}
\label{III}
\begin{figure}[htb]
\begin{center}
\psfig{figure=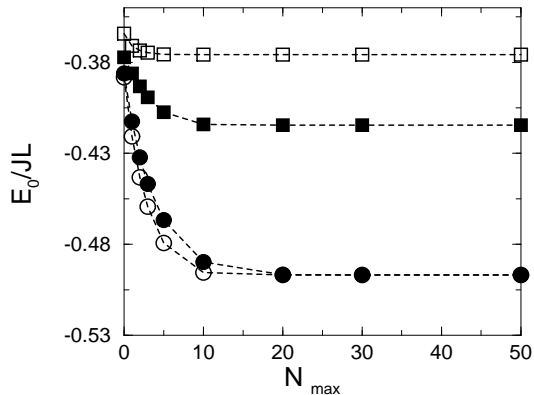,width=7cm,angle=0}
\end{center}
\caption{Convergence of the energy per site of the lowest singlet 
($\circ,\bullet$) and triplet ($\square,\blacksquare$) 
states in units of J as 
a function
of the maximum number of phonons $N_{\mathrm{max}}$. 
Parameters are $\alpha=0.36$, $g=0.5$
and $\Omega=0.3J$. Open (filled) symbols correspond to $L=12$ ($L=20$) sites.}
\label{conv}
\end{figure}

\begin{figure}[htb]
\begin{center}
\psfig{figure=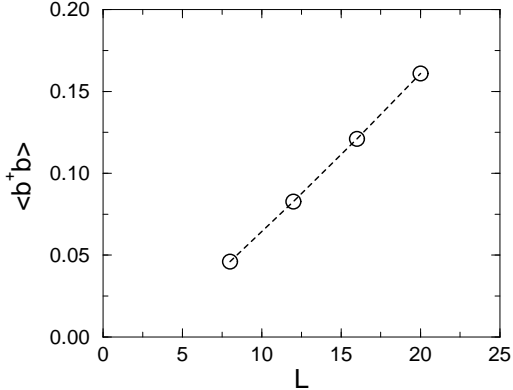,width=7cm,angle=0}
\end{center}
\caption{Dependence of the mean occupation number on the length of the chain
$L$ for $g=0.109$ and $\Omega=0.3J$.}
\label{occ1}
\end{figure}

\begin{figure}[htb]
\begin{center}
\psfig{figure=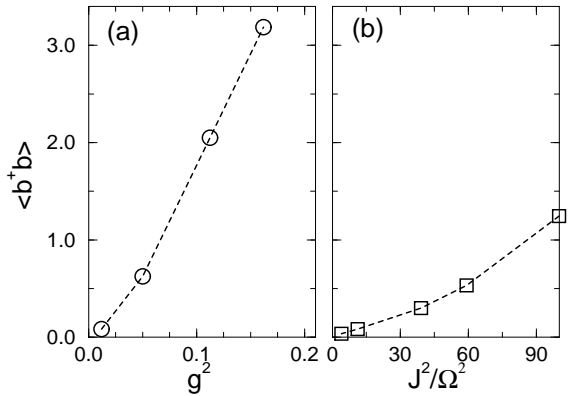,width=7cm,angle=0}
\end{center}
\caption{Mean occupation number calculated on a 
$L=12$ site chain versus $g^2$ for $\Omega/J=0.3$ (a)
and versus $(J/\Omega)^{2}$ for $g=0.109$ (b).
}
\label{occ2}
\end{figure}

Let us now deal first with the numerical treatment of (\ref{hamildyn}). 
The total Hilbert space can be written 
as the tensorial product of the space of the spin configurations 
(to which the symmetry group of the problem is applied) times the 
phononic space. 
However, strictly speaking, the Hilbert space associated 
to the phonons is infinite even for a chain of finite length. Indeed, 
the natural basis $\{|n\rangle\}$ is
simply defined by the unlimited occupation number n of the $k=\pi$ phonon
mode, $|n\rangle =1/\sqrt{n!}\,(b_\pi^\dagger)^n |0\rangle$. Such 
a difficulty can 
nevertheless be easily handled in an exact diagonalization 
treatment~\cite{poilblanc2}. The solution is to truncate the phononic 
space so that the occupation number is  smaller than a fixed 
$N_{\mathrm{max}}$ which has to be chosen in an appropriate 
way. 
Clearly, if $N_{\mathrm{max}}$ is, let us say, an 
order of magnitude larger than the
exact mean occupation number $\langle b_\pi^\dagger b^{\phantom
\dagger}_\pi\rangle$ 
the truncation procedure will not affect the 
accuracy of the results which can then be considered as basically exact.
This can be seen in Fig.~\ref{conv} showing the energy per site of the ground
state in the spin 0 and 1 sectors 
(the value of $\alpha$ corresponds to the case of CuGeO$_3$ 
and a coupling constant $g=0.5$ is used) for chains of length $L=12$ and $20$
plotted as a function of $N_{\mathrm{max}}$.
Typically, the mean occupation number is smaller than 3 as in 
Figs.~\ref{occ1} and \ref{occ2} (and in fact even smaller than $\sim 0.5$ for 
more realistic parameters) and the energy has converged for 
$N_{\mathrm{max}}\sim 30$. 
Fig.~\ref{conv} proves that the truncation procedure is very well 
controlled even for rather (unphysically) large coupling
constants like $g=0.5$. The results reported in the rest of this paper 
are then obtained with a sufficiently large value of $N_{\mathrm{max}}$
and the preliminary studies of the convergence of the results with increasing 
$N_{\mathrm{max}}$, although not mentioned each time, have been performed 
for each choice of the parameters of the model.

It is interesting to study the dependence of the mean occupation number on
the three parameters of the problem (length of the chain $L$,
coupling constant $g$ and frequency of the phonons $\Omega$) since, first,
this number directly determines the practical value of $N_{\mathrm{max}}$
to be chosen and, secondly, it provides some physical 
understanding. Fig.~\ref{conv} already suggests 
that the mean occupation number increases with the length of the chains. 
To investigate this effect in more details,
the mean occupation number $\langle\Psi_0|b^{\dagger}_{\pi}
b_{\pi}|\Psi_0\rangle$ in the ground state $|\Psi_0\rangle$
is plotted in Fig.~\ref{occ1} as a function of $L$.
Clearly, $\langle\Psi_0|b^{\dagger}_{\pi}b_{\pi}|\Psi_0\rangle$ 
(as well as the value required for $N_{\mathrm{max}}$) grows linearly 
with the chain length $L$. In fact, this effect is directly connected 
to the breaking of the lattice symmetry as can be seen very easily 
from a very simplified version of Hamiltonian~(\ref{hamildyn}).
In a symmetry broken state, an effective (approximate) phononic Hamiltonian 
$H_{ph}$ can be constructed by taking MF values for the 
spin operators. Assuming that 
$\sum_i(-1)^i \langle \vec{S}_i.\vec{S}_{i+1}\rangle_{MF}$ 
(dimer order parameter)
varies linearly with $L$ one then gets 
$H_{ph}=Ag\sqrt{L}(b_{\pi}+b^{\dagger}_{\pi})
+\Omega\ b^{\dagger}_{\pi}b_{\pi}$ ($A$ is an
undetermined constant). In this approximation, 
$\langle b^{\dagger}_{\pi}b_{\pi}
\rangle =A^2g^2L/\Omega^2$ grows linearly with the length of the chain.
In addition, this simple argument also suggests that the occupation number 
of the $\pi$ mode scales like the square of the dimensionless coupling $g$ and
like the inverse square of the phonon frequency.
These intuitive behaviors are indeed well followed as can be seen 
in Fig.~\ref{occ2} in a large range of parameters.

One also observes in Fig.~\ref{conv}
that the singlet ground state energy is almost converged for
$L=20$ (the values of the energies for
$L=12$ and $L=20$ at large $N_{\mathrm{max}}$ are indistinguishable) 
while finite size effects are still large for the triplet energy 
because of the existence of a continuum
of states above the first triplet excitation.
In the next Section, we investigate carefully the 
convergence of various physical quantities with respect to the system size.
We show that an accurate finite size analysis 
can be performed to obtain extrapolations to the thermodynamic limit. 

\section{Finite size scaling analysis}
\label{IV}
\begin{figure}[htb]
\begin{center}
\psfig{figure=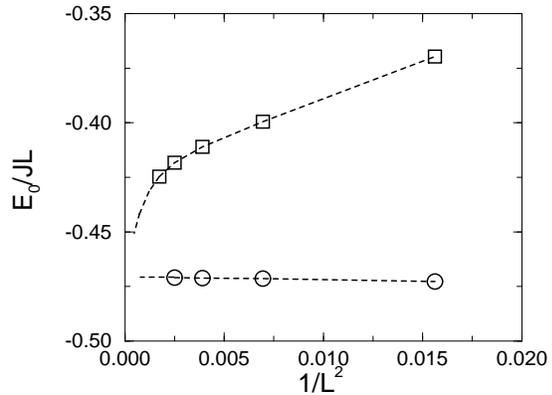,width=7cm,angle=0}
\end{center}
\caption{Convergence of the energy per site 
of the ground state of spin 0 
($\circ$) and 1 ($\square$) in units of J as 
a function of the inverse of the square 
length of the chain $1/L^2$ for $\alpha=0.15$, $g=0.45$
and $\Omega=0.3J$.}
\label{scaling}
\end{figure}

Firstly, we focus on the size dependence of 
the energy {\it per site} of the singlet ground
state and of the lowest triplet state which are expected to converge to 
the same value in the thermodynamic limit.
Typically, we use chains of length $L=8$, $12$, $16$, $20$ and $24$ sites.
Data are shown in Fig.~\ref{scaling} for $\alpha=0.15$, $g=0.45$ and
$\Omega=0.3J$. 
The ground state energy per site varies roughly like $1/L^2$. 
This behavior is predicted for gapless 1D chains obeying conformal 
invariance~\cite{frahm} but seems to be 
still valid here in spite of the presence of a spin gap (see later).
This already suggests that, for such parameters, the system sizes are still 
comparable to the spin correlation length but not much larger.
The behavior of the triplet energy is more involved. 
An approximate $\frac{1}{L}$ dependence is expected (giving a square 
root singularity in the $1/L^2$ units of Fig.~\ref{scaling}) if there is 
a finite spin gap $\Delta$ (defined by the $L\rightarrow\infty$ 
extrapolation of the difference $\Delta(L)=E_{0}(S=1,L)-E_{0}(S=0,L)$
of the {\it total} energies of the lowest states of the
singlet and triplet spin sectors). 
Such a behavior seems indeed to be observed in Fig.~\ref{scaling}.

Let us now examine in details the behavior of the spin gap $\Delta(L)$ 
versus $L$ to extract values in the thermodynamic limit. 
Requiring that the extrapolated ratio $\Delta/J$ of the 
model~(\ref{hamildyn}) is equal to the 
observed experimental value will lead to some constraint on the
model parameters $\Omega$ and $g$.  
Our procedure can be summarised in three steps; (i) a controlled truncation
procedure of the phononic Hilbert space for a large set of parameters $g$, 
$\Omega$ and system sizes $L$, (ii) a finite size scaling analysis in order 
to accurately determine the spin gap as a function of $g$ and 
$\Omega$; (iii) a determination of the relation to be followed by 
the parameters $g$ and $\Omega$ in order that the calculated ratio $\Delta/J$ 
equals the actual experimental ratio (see Section~\ref{V}).

\begin{figure}[hbt]
\begin{center}
\psfig{figure=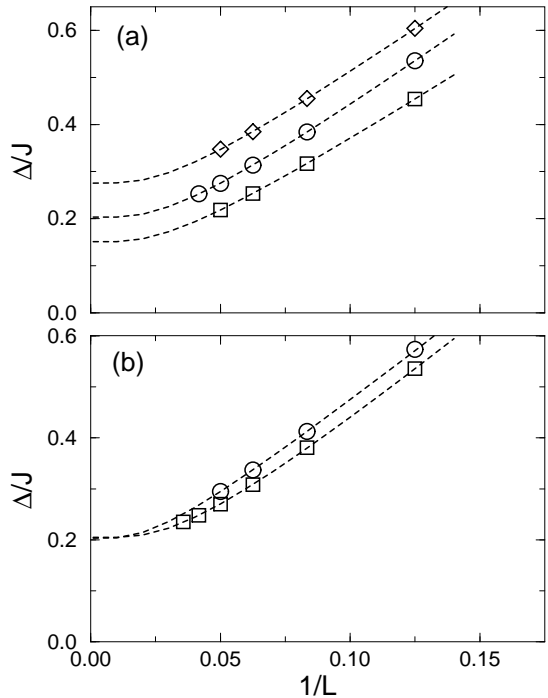,width=7cm,angle=0}
\end{center}
\caption{(a) Spin gap $\Delta$ in units of $J$ as a function of
the inverse of the 
length of the chain $1/L$ for $\alpha=0.15$, $g=0.22$, $\Omega=0.3J$
($\circ$), $\alpha=0.36$, $g=0.089$, $\Omega=0.3J$ ($\square$) and
$\alpha=0$, $g=0.40$, $\Omega=0.5J$ ($\lozenge$).
(b) Comparison between the behaviors $\Delta/J$ vs $1/L$ obtained
within the dynamical model~(\ref{hamildyn}) for $\alpha=0$, $g=0.275$, 
and $\Omega=0.3J$ ($\circ$) and within the static model~(\ref{hamilstat}) 
for $\alpha=0$, $\delta=0.05$ ($\square$).}
\label{scalgap}
\end{figure}

We first consider the scaling behavior of the spin gap.
We have found that it scales accurately
according to the law,~\cite{barber,bouzerar}
\begin{equation}
\Delta(L)=\Delta+\frac{A}{L}\exp(-\frac{L}{L_0}),
\end{equation}
where  $L_0$ is a typical length scale. In general $L_0$ is 
of the order of the magnetic correlation 
length characterizing the decay of the equal time 
spin-spin correlation in real space. As seen later, values of $L_0$ 
are typically 20 lattice units (l.u.) for parameters corresponding to 
CuGeO$_3$ and 30 l.u. for NaV$_2$O$_5$.  
Therefore, with chains lengths up to 24 sites,
finite size effects are still important and an accurate extrapolation 
is necessary. This scaling is illustrated 
for $\alpha=0.15<\alpha_c$, $g=0.22$, $\Omega=0.3J$ ($\circ$),
for $\alpha=0.36>\alpha_c$ (CuGeO$_3$-like case), $g=0.089$, $\Omega=0.3J$ 
($\square$) 
and for $\alpha=0$ (NaV$_2$O$_5$-like case),
$g=0.40$, $\Omega=0.5J$ 
($\lozenge$) in Fig.~\ref{scalgap}(a).
A spin gap opens for all $\alpha$ if $g>0$. This is similar to the 
mean-field treatment where the order parameter $\delta\neq 0$
leads to the symmetry breaking and thus to the opening of a spin gap. 

In Fig.~\ref{scalgap}(b)
we compare, in the case of
NaV$_2$O$_5$ (i.e. $\alpha=0$), the scaling of the spin gaps 
calculated using the dynamical model~(\ref{hamildyn}) with $g=0.275$,
$\Omega=0.3J$ ($\circ$) on one hand and the static 
model~(\ref{hamilstat}) with $\delta=0.05$ ($\square$) on the other 
hand~\cite{note_static}. These values of the parameters 
have been chosen in order to obtain the same extrapolated spin gap. 
Although the spin gaps are equal,
the two models exhibit slightly different scaling behaviors 
($L_0\simeq 30$ for the dynamical model and $L_0\simeq18$ for the 
static one~\cite{augier}).

\begin{figure}[htb]
\begin{center}
\psfig{figure=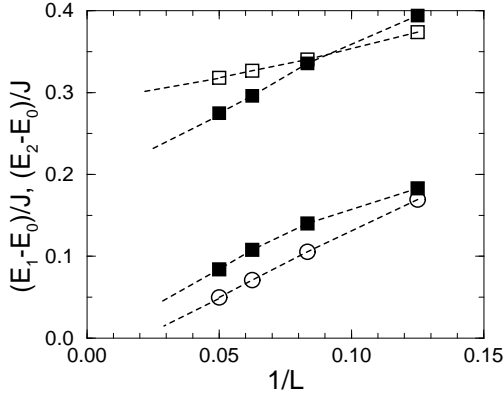,width=7cm,angle=0}
\end{center}
\caption{Energy differences $(E_1(S=0)-E_0(S=0))/J$ ($\circ,\bullet$)
and $(E_2(S=0)-E_0(S=0))/J$ ($\square,\blacksquare$) 
as a function of $1/L$. Open and filled symbols correspond to 
$\alpha=0.36$, $g=0.109$ and $\Omega=0.3J$ and to 
$\alpha=0$, $g=0.270$ and $\Omega=0.3J$, respectively.}
\label{deg}
\end{figure}

At this stage, it is interesting to better understand how
in the the dynamical model~(\ref{hamildyn}) the opening of the spin gap
is connected to the discrete symmetry breaking (as can be seen e.g. in
X-rays scattering). The first signature of this phenomenon 
is the degeneracy of the ground state which is expected in the 
thermodynamic limit.
We have therefore studied the behavior with system size of 
the energies $E_p(S=0)$, $p=0,1,2$, of the three lowest singlet states.
The energy differences $E_1(S=0)-E_0(S=0)$ (circles) and $E_2(S=0)-E_0(S=0)$
(squares) are plotted in Fig.~\ref{deg}, in the case $\Omega=0.3J$, 
as a function of the inverse 
length of the chain $1/L$ for $\alpha=0.36$ (open symbols) and for 
$\alpha=0$ (filled symbols). The values of the coupling $g$ are chosen 
here in such a way to reproduce the experimental spin gaps of the  
CuGeO$_3$ (open symbols) and NaV$_2$O$_5$ (filled symbols)
materials (see Section V). 
The results show very convincingly that the singlet ground state 
is indeed two-fold degenerate in the thermodynamic limit while a finite
gap for singlet excitations appears above~\cite{note}.
It is important to notice that the quantum numbers associated to
the translation symmetry are different for the two lowest singlet states 
which correspond to momenta $k=0$ and $k=\pi$.
Hence, mixing of these two states leads to a doubling of the unit cell.

\begin{figure}[htb]
\begin{center}
\psfig{figure=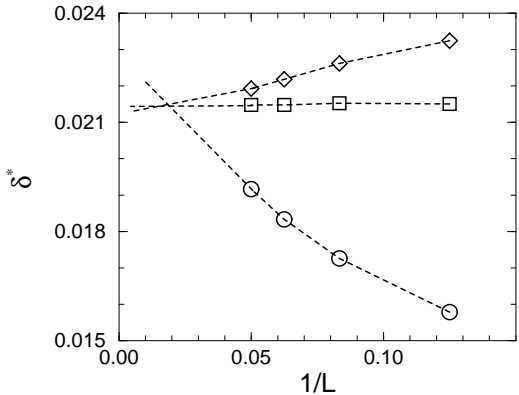,width=7cm,angle=0}
\end{center}
\caption{Order parameter $\delta^*$ as a function of the inverse of the length
of the chain $1/L$ for $\Omega=0.1J$ and $g=0.062$ ($\circ$),
$\Omega=0.3J$ and $g=0.109$ ($\square)$ and $\Omega=0.5J$ and $g=0.141$ 
($\lozenge$) (see text regarding the choice of parameters).}
\label{depl}
\end{figure}

The lattice dimerization can be quantitatively measured by the order
parameter $\delta^*=\frac{g}{\sqrt{L}}
\langle(b^{\phantom\dagger}_{\pi}+b_{\pi}^{\dagger})^2\rangle^{1/2}$ (the expectation
value $\langle b^{\phantom\dagger}
_{\pi}+b_{\pi}^{\dagger}\rangle$ vanishes because
small tunnelling between the two degenerate dimer states always 
exists in a finite chain).
$\delta^*$ as a function of the inverse length of the
chain is plotted in Fig.~\ref{depl} for various pairs of
parameters $(\Omega,g)$ (see caption) chosen in such a way that the 
spin gap is constant (in fact adjusted to the actual spin 
gap of CuGeO$_3$ as described in Section V). 
Extrapolated values of the dimerization $\delta^*$ for different 
phonon frequencies are in fact quite close, at least in the 
range $0.1\le \Omega\le 0.5$. The dimerization $\delta^*$ 
seems then to be only determined by the magnitude of the
spin gap. 
The fact that $\delta^*$, at fixed extrapolated spin gap, is independent 
of the frequency $\Omega$ is consistent with the proof by
Brandt and Leschke~\cite{lieb} that the thermodynamic properties of the 
dynamical model~(\ref{hamildyn}) and of the static model~(\ref{hamilstat}) 
are identical. However,
it is interesting to notice that the value obtained here ($\sim 0.022$) 
is significantly larger than the value ($\sim 0.014$) needed in the MF 
approximation to produce the same gap.  
The difference between these two values can be simply attributed to the
zero point motion of the harmonic mode which is included only 
in (\ref{hamildyn}).

\section{Comparison with experiment}
\label{V}
\begin{figure}[htb]
\begin{center}
\psfig{figure=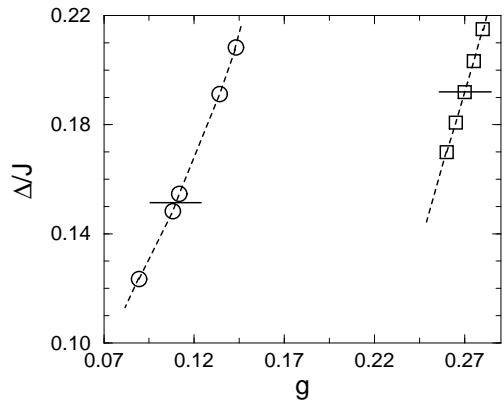,width=7cm,angle=0}
\end{center}
\caption{Spin gap $\Delta/J$ (in units of $J$) as a function of
the magneto-elastic coupling $g$ for $\alpha=0.36$, $\Omega=0.3J$
($\circ$) and $\alpha=0$, $\Omega=0.3J$ ($\square$). Horizontal marks
indicate actual experimental spin gap values.}
\label{gap}
\end{figure}

The systematic finite size scaling described above has been performed 
for a large set of parameters $\Omega/J$ and $g$.
For simplicity, let us first assume $\Omega/J=0.3$.
The behavior of $\Delta(\Omega/J=0.3,g)/J$ versus $g$ is plotted in 
Fig.~\ref{gap} for a large frustration $\alpha=0.36$ corresponding
to the case of CuGeO$_3$ ($\circ$) and for a non frustrated chain 
corresponding to the case of NaV$_2$O$_5$ ($\square$).
Quite generally, the spin gap grows with the coupling constant $g$
as expected. Indeed, a larger coupling to the lattice produces a larger
dimerization and then, indirectly, a larger spin gap.

\begin{figure}[htb]
\begin{center}
\psfig{figure=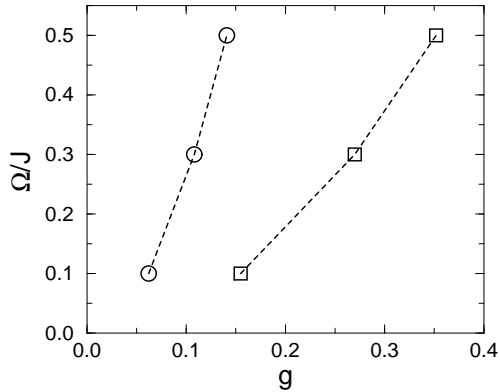,width=7cm,angle=0}
\end{center}
\caption{Frequency $\Omega$ in units of $J$ as a function of
the magneto-elastic coupling $g$ insuring a constant spin gap (see text) for
$\alpha=0.36$ ($\circ$) and $\alpha=0$ ($\square$).}
\label{freq}
\end{figure}

The actual physical value of the ratio $\Omega/J$ is, to the best
of our knowledge, difficult to obtain from experiment. Therefore,
we shall not here restrict to any specific value of  $\Omega/J$ but rather
consider a wide range $0.1\le \Omega/J\le 0.5$.
However, for each value of $\Omega$, the dimensionless coupling constant
$g(\Omega)$ can be determined by enforcing that the extrapolated spin
gap ratio $\Delta(\Omega,g)/J$ equals the experimentally observed 
gap. The procedure is shown in Fig.~\ref{gap} for $\Omega=0.3J$
and $\alpha=0$ (NaV$_2$O$_5$) and $\alpha=0.36$ (CuGeO$_3$).
The small horizontal marks correspond to the actual experimental 
gaps, i.e. $\Delta/J\simeq0.151$ and $\Delta/J\simeq0.193$ 
for CuGeO$_3$ and NaV$_2$O$_5$, respectively.
We then obtain $g(\Omega=0.3)\simeq0.109$ for CuGeO$_3$ 
and $g(\Omega=0.3)=0.270$ for NaV$_2$O$_5$.
The same method was performed for two other values of the
frequency, $\Omega=0.1J$ and $\Omega=0.5J$. 
A relation is then obtained between $\Omega$ and $g$ for the
two values of the frustration parameter $\alpha=0$ and $\alpha=0.36$.
This is illustrated in Fig.~\ref{freq}. We find that $\Omega$ 
has to vary roughly like $g^2$ in order that the spin gap is constant.
Naively, one indeed expects that softer ({\it i.e.} with smaller $\Omega$) 
phonon modes are more effective to break the lattice symmetry. 
So, if one requires the spin gap to be constant, this effect has to be 
compensated by a smaller coupling $g$.

In Fig.~\ref{freq} we observe that
the coupling constant $g(\Omega)$ is roughly $2.5-3$ 
times smaller for CuGeO$_3$ than for NaV$_2$O$_5$ 
although the ratio of their spin gaps is only $1.5$. 
This is an interesting consequence of the large 
frustration in CuGeO$_3$. Indeed, a large $\alpha$ opens 
alone a (quite small) spin gap and, more importantly, amplifies the 
effect of the spin-phonon coupling. 
This effect is even more drastic in the static model~(\ref{hamilstat}) 
where the dimerizations $\delta=0.014$ (CuGeO$_3$) 
and $\delta=0.048$ (NaV$_2$O$_5$)
have a ratio of about $4$~\cite{augier}.

The model~(\ref{hamildyn}) seems to describe accurately the
spin-Peierls transition. Theoretical parameters have been 
deduced from experiment and the ground state properties of the 
spin-Peierls phase have been established.
We have provided evidences in favour of the dynamical breaking of the 
lattice periodicity with the simultaneous opening of the spin gap. 
Next, we shall study the dynamical properties of this model.

\section{Dynamical properties}
\label{VI}
INS is a powerful experiment 
probing the momentum-dependence of the spin dynamics. 
INS has been performed on CuGeO$_3$ single crystals~\cite{regnault,nishi}  
and on NaV$_2$O$_5$ powders~\cite{fujii}. It provides a
direct measure of the dynamical spin-spin structure factor,
\begin{equation}
S_{zz}(q,\omega)=\sum_n|\langle \Psi_n|S_z(q)|\Psi_0\rangle|^2
\delta(\omega-E_n+E_0),
\label{struc}
\end{equation}
where $|\Psi_0\rangle$ is the (singlet) ground state of energy $E_0$
and the sum is performed on all triplet excited states 
$|\Psi_n\rangle$ (of energy $E_n$). $S_z(q)$ is normalised as 
$1/\sqrt{L}\sum_j \exp(iqj) S_j^Z$. 

\begin{figure}[htb]
\begin{center}
\psfig{figure=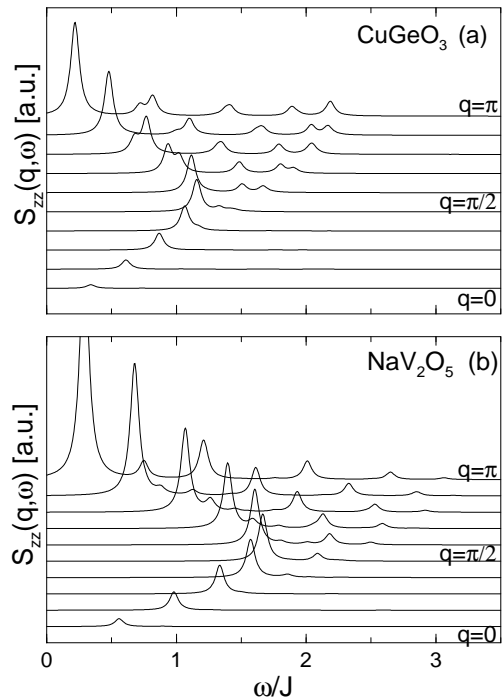,width=7cm,angle=0}
\end{center}
\caption{$S_{zz}(q,\omega)$ as a function of $\omega/J$ calculated
on a 20 site chain for 
$q=n\frac{\pi}{10}$, $n=0,\cdots,10$; (a) CuGeO$_3$ 
parameters, $\alpha=0.36$, $g=0.109$, $\Omega=0.3J$;
(b) NaV$_2$O$_5$ 
parameters, $\alpha=0$, $g=0.270$, $\Omega=0.3J$.
A broadening of the $\delta$-functions $\varepsilon=0.04J$ was used.}
\label{szzqw}
\end{figure}

The INS spectrum can be easily computed 
by exact diagonalization techniques~\cite{poilblanc2}. 
Results on a 20 site chain are shown in Fig.~\ref{szzqw}(a) for CuGeO$_3$
and in Fig.~\ref{szzqw}(b) for NaV$_2$O$_5$ with a frequency
$\Omega=0.3J$. In both cases we observe a well defined q-dependent low energy
structure as for the static model~(\ref{hamilstat}). 
Its bandwidth (i.e. the energy at the maximum of the dispersion at
$q=\pi/2$) is typically  $\omega_{max}\sim1.1J$ for
CuGeO$_3$ and $\omega_{max}\sim1.6J$ for NaV$_2$O$_5$. This second
value is very close to the DesCloizeaux-Pearson 
value of $\pi/2$~\cite{descloiseaux} of the Heisenberg chain
in contrast to the case of CuGeO$_3$ which exhibits a large frustration.
The ratio $\omega_{max}/J$ could therefore be considered as 
an additional accurate measure of the amount of frustration 
within the chain since the parameter $\alpha$ alone 
determine approximately $\omega_{max}/J$. 
It is interesting to notice also that, in the case of 
a frustrated chain (CuGeO$_3$),
the upper limit of the continuum seems to be better defined.

At low energy, the dimerization gap leads to major differences with respect to 
the Heisenberg chain. First, there is no intensity for $\omega<\Delta$.
Secondly, the magnon branch is well separated from the continuum by
a finite gap (see below) so that the magnon excitation can be interpreted as a 
spinon-spinon bound state~\cite{uhrig}. This bound state was also found in
the static model~(\ref{hamilstat}) for CuGeO$_3$~\cite
{fledder,poilblanc} and NaV$_2$O$_5$~\cite{augier}.

\begin{figure}[htb]
\begin{center}
\psfig{figure=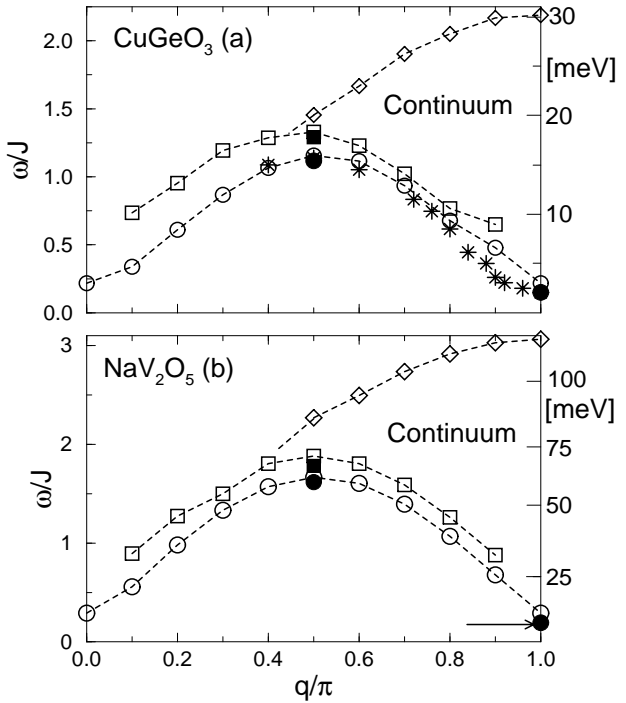,width=7cm,angle=0}
\end{center}
\caption{Momentum dependence of the first ($\circ$), 
second excitation ($\square$) and upper limit of
the continuum ($\lozenge$) on a 20 site chain
for the dynamical model~(\ref{hamildyn}). Filled symbols 
represent extrapolations
to infinite size (first ($\bullet$) and second ($\blacksquare$)
excitations).
(a) CuGeO$_3$ parameters ($\alpha=0.36$, $g=0.109$,
$\Omega=0.3J$). Experimental values ($\ast$) are taken from
from Ref.~\protect\cite{regnault}. Units on the right are in 
meV assuming $J=160K$ ($13.8$~meV). (b) NaV$_2$O$_5$ 
parameters ($\alpha=0$, $g=0.270$,
$\Omega=0.3J$). Units on the right are in meV assuming $J=440K$
($37.9$~meV). The
arrow indicates the experimental value of the $q=\pi$ spin gap.}
\label{disp}
\end{figure}

The dispersion relations of the magnon branch ($\circ$),  
the second excitation ($\square$) and
the upper limit of the continuum ($\lozenge$) in the
dynamical model~(\ref{hamildyn}) are plotted in 
Fig.~\ref{disp}(a) for CuGeO$_3$. The `$\ast$' symbols correspond to
experimental results from Ref.~\cite{regnault} and filled symbols
correspond to infinite size extrapolations
at momenta $q=\pi/2$ and $q=\pi$. Similar dispersion relations 
are shown in Fig.~\ref{disp}(b) for  
NaV$_2$O$_5$ and the position of
the experimental $q=\pi$ spin gap~\cite{fujii} is indicated by an arrow.
Note that we have explicitly checked that the magnon branch is 
well separated from the continuum. A finite size scaling analysis 
of the energies of the two lowest triplet states ($\bullet$,$\blacksquare$)
is indeed possible at momentum $q=\pi/2$. Figs.~\ref{disp}(a-b) clearly
show that there is a finite gap between the first branch and the continuum
as in the static model. It is consistent with the fact that the 
continuum corresponds to solitonic spin-1/2 excitations (or spinons) and that 
solitons and antisolitons can bind in pairs with 
momenta close to $q=\pi/2$~\cite{uhrig}.
Such a dougle gap feature was indeed observed experimentally~\cite{ain}.

It is important to notice that the dispersion relation is not symmetric 
with respect to
$q=\pi/2$ in contrast to the case of a static dimerization. In fact, such
a symmetry in the energy spectrum is due to the Bragg scattering
resulting from the doubling of the unit cell. Since 
the dimerization appears only as a true phase transition
in model~(\ref{hamildyn}), we expect that the symmetry of the spectrum
with respect to $\pi/2$ will only become exact in the thermodynamic
limit. In the
case of CuGeO$_3$, our results are in very good agreement with
INS experiments although finite size effects are still important. 
In fact the agreement improves with increasing system 
size since the calculated magnon branch for $q>\pi/2$ shifts slightly to
lower energy when $L$ grows (in order to be
symmetric with the $q<\pi/2$ part). 
Note also that energy scales are four times larger for
NaV$_2$O$_5$ than for CuGeO$_3$ which could restrict 
INS experiments on NaV$_2$O$_5$ to
low energy regions of the spectrum in the vicinity of $q=\pi$.

\begin{figure}[htb]
\begin{center}
\psfig{figure=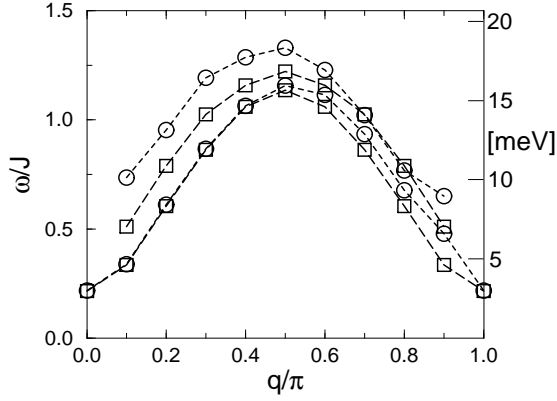,width=7cm,angle=0}
\end{center}
\caption{Momentum dependence of the two lowest triplet excitation
energies in CuGeO$_3$ calculated on a 20 site chain
for (i) the dynamical model~(\ref{hamildyn}) ($\alpha=0.36$, $g=0.109$,
$\Omega=0.3J$) ($\circ$) and (ii) the static model~(\ref{hamilstat}) 
($\alpha=0.36$, $\delta=0.014$) ($\square$). Units on the right are meV
assuming that $J=160K$ ($13.8$~meV).}
\label{dispcomp}
\end{figure}

It is interesting to compare results for the spin dynamics 
obtained within the dynamical model to the ones obtained
within the static model.
Fig.~\ref{dispcomp} shows the lowest triplet magnon branches
and the next triplet excitations (in fact lower limits of the 
$S=1$ continuum) for parameters suitable for  CuGeO$_3$. 
We do not explicitly show the comparison of the upper limits of
the continua since the two curves obtained within the two models
are almost indistinguishable. This is not surprising because 
higher energy excitations are only determined by the magnitude of the
frustration and the coupling to the lattice plays a minor role here. 
At lower energy, the magnon branches of the two models look also very
similar for $q<\pi/2$ but some differences appear for $q>\pi/2$ since,
as explained before, the dispersion is not symmetric with respect 
to $\pi/2$ in the dynamical model. This is simply due to larger finite 
size effects~\cite{note2} 
occurring in model (\ref{hamildyn}) related to
the fact that the lattice periodicity is 
only {\it spontaneously} broken.
Once such finite size effects are taken into account 
we can safely conclude that the dispersions of the magnon branches of the
two models in the thermodynamic limit are very close.
Similarly, the discrepancies seen between the positions of the lower 
limits of the
continua of triplet excitations are not relevant. Indeed, a detailed 
finite size scaling analysis at e.g. $q=\pi/2$ reveals that the 
position of the two lower limits
are in fact quite close ($1.117J$ for (\ref{hamilstat}) to be compared 
to $1.118J$ for (\ref{hamildyn})).
An exactly similar comparison can be done for 
NaV$_2$O$_5$ (not shown). 
 
\begin{figure}[htb]
\begin{center}
\psfig{figure=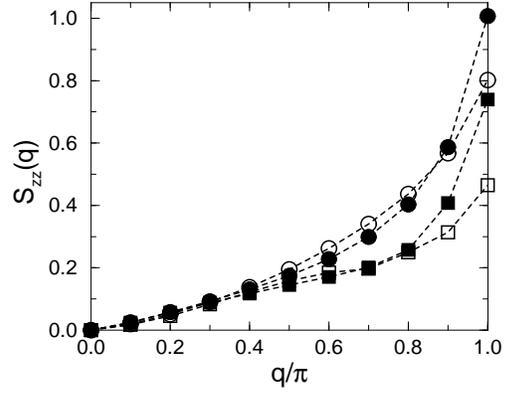,width=7cm,angle=0}
\end{center}
\caption{Static factor structure for CuGeO$_3$ ($\alpha=0.36$, $g=0.109$, 
$\Omega=0.3J$) ($\circ$) 
and NaV$_2$O$_5$ ($\alpha=0$, $g=0.270$, $\Omega=0.3J$) 
($\bullet$) calculated on a 20 site chain. Squares correspond to
the weight of the lowest peak 
for CuGeO$_3$ ($\square$) and for NaV$_2$O$_5$
($\blacksquare$).}
\label{stat}
\end{figure}

The spin static structure factor,
$$S_{zz}(q)=\int d\omega S_{zz}(q,\omega)\ ,$$
which can be obtained in INS by integrating the spectrum over energy
is plotted in Fig.~\ref{stat} for CuGeO$_3$ ($\alpha=0.36$, $g=0.109$, 
$\Omega=0.3J$) ($\circ$)
and NaV$_2$O$_5$ ($\bullet$)
($\alpha=0$, $g=0.270$, $\Omega=0.3J$) 
for a 20 site chain. It is peaked near $q=\pi$ as a result
of strong short range AF correlations. Indeed the width of the 
peak at $q=\pi$ is directly related to the inverse 
magnetic correlation length. 
Note however that $S_{zz}(\pi)$ is slightly suppressed in CuGeO$_3$
compared to NaV$_2$O$_5$ because of the interchain frustration. In any case,
the results are very similar to those obtained
with the static dimerized model. The relative weights of the magnon peak
in $S_{zz}(q,\omega)$ are also shown for CuGeO$_3$ ($\square$) and
NaV$_2$O$_5$ ($\blacksquare$).
Their behaviors versus $q$ suggest that working in
a range of momenta around 
$q=0.8\pi$ might be more appropriate experimentally in order to
have clearer evidences for the continuum.

\section{Conclusions}
In order to describe one dimensional spin-Peierls compounds, 
a magneto-elastic (i.e. spin-phonon) coupling has been considered
and is shown to be responsible for a dynamical and spontaneous breaking of the
lattice periodicity followed simultaneously by the opening of a 
spin gap. 
The resulting symmetry-broken ground state is consistent with the
existence of a frozen dimerization such as the one obtained 
in a mean-field treatment of the coupling to the lattice. 
We have used exact diagonalization techniques to 
calculate static and dynamical properties of this model.
Controlled truncation procedures have been applied to the bosonic
Hilbert space of the Hamiltonian. 
By using a finite size scaling analysis, we have compared various 
physical quantities to the experimental ones 
(in the case of CuGeO$_3$ and NaV$_2$O$_5$) and we have determined 
a range of suitable parameters for the model. 
We predict that the spin-phonon coupling is 2 or 3 times
larger in NaV$_2$O$_5$ than in CuGeO$_3$.
The INS spectrum calculated within this model is found to be 
qualitatively similar to the one obtained in the static model with
a finite gap separating the magnon branch from the continuum of 
triplet excitations above. 

\bigskip
D.A. acknowledges useful discussions with
M.~Albrecht and S.~Capponi. We thank L.~P. Regnault for
communicating to us the data of Ref.~\cite{regnault},
J.~Riera for valuable comments and IDRIS (Orsay) 
for allocation of CPU time on the C94 and C98 CRAY supercomputers.

After completion of this work, we learnt of a related work by 
A.~W.~Sandvik {\em et al.} (cond-mat/9706046) using a
Quantum Monte Carlo approach.


\begin{references}
\bibitem{notte} The equivalence has been rigorously established 
by Ref.~\protect\cite{lieb} for the Fr\"ohlich Hamiltonian and 
should be also exact for model (2) which includes correlations between 
electrons.
\bibitem{note_static} For the static model~(\protect\ref{hamilstat}),
chains with up to $L=28$ sites were considered.
\bibitem{note} The second singlet double-magnon excitation energy is roughly 
twice larger than the first triplet excitation 
(for more details see e.g.~\protect\cite{augier}). However, singlet 
phononic excitations may lie below the continuum.
\bibitem{note2} As can be seen e.g. in the finite size 
extrapolations of the triplet excitations energies at $q=\pi/2$.
In contrast, in the static model~(\protect\ref{hamilstat}),
the computed data~\protect\cite{poilblanc,augier} are almost indistinguishable
from the extrapolated values.
\end{references}
\end{document}